\documentstyle[12pt]{article}
\newlength{\pubnumber} 

\catcode`\@=11
\@addtoreset{equation}{section}
\def\section{\@startsection{section}{1}{\z@}{3.5ex plus 1ex minus .2ex}
 {2.3ex plus .2ex}{\large\bf}}
\def\subsection{\@startsection{subsection}{2}{\z@}{2.3ex plus .2ex}
 {2.3ex plus .2ex}{\bf}}

\begin{document}

\begin{titlepage}
\samepage{
\setcounter{page}{1}
\rightline{OUTP--00--46P}
\rightline{\tt hep-th/0011006}
\rightline{November 2000}
\vfill
\begin{center}
 {\Large \bf Proton Stability and Superstring Z$^\prime$\\}
\vfill
\vspace{.25in}
 {\large Alon E. Faraggi\\}
\vspace{.25in}
 {\it  Theoretical Physics Department,\\
              University of Oxford, Oxford, OX1 3NP, United Kingdom\\}
\vspace{.05in}
\end{center}
\vfill
\begin{abstract}
  {\rm
Recently it was argued that proton lifetime limits impose that the 
scale of quantum gravity must be above $10^{16}${\rm GeV}. By studying
the proton stability in the context of realistic heterotic--string models,
I propose that proton longevity necessitates the existence of an 
additional $U(1)_{Z^\prime}$ symmetry, which is of non--GUT origin
and remains unbroken down to intermediate, or low, energies. 
It is shown that the realistic string models frequently give rise to 
$U(1)$ symmetries, which suppress the proton decay mediating operators,
with, or without, the possibility of R--parity violation. By studying
the F-- and D--flat directions, I examine whether the required symmetries
remain unbroken in the supersymmetric vacuum and show that in some examples
they can, whereas in others they cannot.
The proton decay rate is proportional to the $U(1)_{Z^\prime}$
symmetry breaking scale. Imposing the proton lifetime limits I estimate the
$U(1)_{Z^\prime}$ breaking scale and show that if substantial R--parity
violation is present the associated $Z^\prime$ is within reach of 
forthcoming collider experiments.}
\end{abstract}
\vfill
\smallskip}
\end{titlepage}

\setcounter{footnote}{0}

\def\beq{\begin{equation}}
\def\eeq{\end{equation}}
\def\beqn{\begin{eqnarray}}
\def\eeqn{\end{eqnarray}}
\def\Tr{{\rm Tr}\,}
\def\KM{{Ka\v{c}-Moody}}

\def\ie{{\it i.e.}}
\def\etc{{\it etc}}
\def\eg{{\it e.g.}}
\def\half{{\textstyle{1\over 2}}}
\def\third{{\textstyle {1\over3}}}
\def\quarter{{\textstyle {1\over4}}}
\def\m{{\tt -}}
\def\p{{\tt +}}

\def\rep#1{{\bf {#1}}}
\def\slash#1{#1\hskip-6pt/\hskip6pt}
\def\slk{\slash{k}}
\def\GeV{\,{\rm GeV}}
\def\TeV{\,{\rm TeV}}
\def\y{\,{\rm y}}
\def\SM{Standard-Model }
\def\SUSY{supersymmetry }
\def\SSM{supersymmetric standard model}
\def\vev#1{\left\langle #1\right\rangle}
\def\l{\langle}
\def\r{\rangle}

\def\Htw{{\tilde H}}
\def\chibar{{\overline{\chi}}}
\def\qbar{{\overline{q}}}
\def\ibar{{\overline{\imath}}}
\def\jbar{{\overline{\jmath}}}
\def\Hbar{{\overline{H}}}
\def\Qbar{{\overline{Q}}}
\def\abar{{\overline{a}}}
\def\alphabar{{\overline{\alpha}}}
\def\betabar{{\overline{\beta}}}
\def\tautwo{{ \tau_2 }}
\def\calF{{\cal F}}
\def\calP{{\cal P}}
\def\calN{{\cal N}}
\def\smallmatrix#1#2#3#4{{ {{#1}~{#2}\choose{#3}~{#4}} }}
\def\bone{{\bf 1}}
\def\V{{\bf V}}
\def\b{{\bf b}}
\def\N{{\bf N}}
\def\bQ{{\bf Q}}
\def\t#1#2{{ \Theta\left\lbrack \matrix{ {#1}\cr {#2}\cr }\right\rbrack }}
\def\C#1#2{{ C\left\lbrack \matrix{ {#1}\cr {#2}\cr }\right\rbrack }}
\def\tp#1#2{{ \Theta'\left\lbrack \matrix{ {#1}\cr {#2}\cr }\right\rbrack }}
\def\tpp#1#2{{ \Theta''\left\lbrack \matrix{ {#1}\cr {#2}\cr }\right\rbrack }}


\def\inbar{\,\vrule height1.5ex width.4pt depth0pt}

\def\IC{\relax\hbox{$\inbar\kern-.3em{\rm C}$}}
\def\IQ{\relax\hbox{$\inbar\kern-.3em{\rm Q}$}}
\def\IR{\relax{\rm I\kern-.18em R}}
 \font\cmss=cmss10 \font\cmsss=cmss10 at 7pt
\def\IZ{\relax\ifmmode\mathchoice
 {\hbox{\cmss Z\kern-.4em Z}}{\hbox{\cmss Z\kern-.4em Z}}
 {\lower.9pt\hbox{\cmsss Z\kern-.4em Z}}
 {\lower1.2pt\hbox{\cmsss Z\kern-.4em Z}}\else{\cmss Z\kern-.4em Z}\fi}

\def\AEF{A.E. Faraggi}
\def\KRD{K.R. Dienes}
\def\JMR{J. March-Russell}
\def\NPB#1#2#3{{\it Nucl.\ Phys.}\/ {\bf B#1} (#2) #3}
\def\PLB#1#2#3{{\it Phys.\ Lett.}\/ {\bf B#1} (#2) #3}
\def\PRD#1#2#3{{\it Phys.\ Rev.}\/ {\bf D#1} (#2) #3}
\def\PRL#1#2#3{{\it Phys.\ Rev.\ Lett.}\/ {\bf #1} (#2) #3}
\def\PRT#1#2#3{{\it Phys.\ Rep.}\/ {\bf#1} (#2) #3}
\def\MODA#1#2#3{{\it Mod.\ Phys.\ Lett.}\/ {\bf A#1} (#2) #3}
\def\IJMP#1#2#3{{\it Int.\ J.\ Mod.\ Phys.}\/ {\bf A#1} (#2) #3}
\def\nuvc#1#2#3{{\it Nuovo Cimento}\/ {\bf #1A} (#2) #3}
\def\etal{{\it et al,\/}\ }

\hyphenation{su-per-sym-met-ric non-su-per-sym-met-ric}
\hyphenation{space-time-super-sym-met-ric}
\hyphenation{mod-u-lar mod-u-lar--in-var-i-ant}


\setcounter{footnote}{0}
\section{Introduction}

The proton longevity is one of the most important guides in attempts
to understand the fundamental origin of the observed gauge and
matter particle spectrum. While the Standard Model does not allow
for the existence of renormalizable operators which can mediate
proton decay, this is not the case in most of its theoretical
extensions. Moreover, even if we assume that the Standard Model
remains unmodified up to the cutoff scale set by quantum gravity,
baryon and lepton number violating operators will in general be induced
at that scale. In fact, recently it was argued, on general grounds,
that proton lifetime limits impose that the cutoff scale must be
above $10^{16}{\rm GeV}$ \cite{kane}. 

The proton longevity problem becomes
especially acute in supersymmetric extensions of the Standard
Model \cite{nilles}, which allow dimension four and five baryon and lepton
number violating operators \cite{WSY}. In the Minimal Supersymmetric 
Standard Model one imposes the existence of a global symmetry,
$R-$parity, which forbids the dangerous dimension four
operators, while the difficulty with the dimension five
operators can only be circumvented if one further assumes
that the relevant Yukawa couplings are sufficiently suppressed. 
However, in general, global symmetries are not preserved in
quantum gravity \cite{Hawking,EHNT}. To satisfy proton lifetime constraints
one must therefore assume the existence of a local discrete
symmetry \cite{lds} or an explicit gauge symmetry. An example of
such a symmetry is the gauged $B-L$ symmetry which forbids
the dimension four proton decay mediating operators of the MSSM.

Realistic superstring models provide a concrete framework 
to study in detail the issue of proton stability in the context
of quantum gravity. Indeed the issue has been examined in the
past by a number of authors \cite{psinsm,ps94,pati}. The avenues
explored range from the existence of matter parity at special
points in the moduli space of specific models, to the emergence
of non--Abelian custodial symmetries in specific compactifications.

The most realistic string models constructed to date are 
the models constructed in the free fermionic formulation \cite{fff}.
This has given rise to a large set of semi--realistic
models \cite{fsu5,fny,pssm,eu,nahe,cus,cfn,cfs},
which differ in their detailed phenomenological 
characteristics, and share an underlying $Z_2\times Z_2$
orbifold structure \cite{foc}. The important achievements include:
the natural emergence of three generations, which is correlated
with the structure of the underlying $Z_2\times Z_2$ orbifold;
The $SO(10)$ embedding of the Standard Model spectrum, yielding
the canonical $SO(10)$ normalization for the weak hypercharge.
Recently, it was further demonstrated that free fermionic construction
also gives rise to models in which the low energy states, which
carry Standard Model charges, consist solely of the spectrum
of the Minimal Supersymmetric Standard Model \cite{cfn}. 
The realistic free fermionic models therefore provide a concrete
and viable framework to study the proton lifetime problem.
In this context
past investigations have examined several possibilities that
may explain the proton longevity. For example, ref. \cite{lepzp}
stipulated the possibility that the $U(1)_{Z^\prime}$ which 
is embedded in $SO(10)$ remains unbroken down to the TeV
scale, and consequently the problematic dimension 4 operators
are adequately suppressed. In ref. \cite{psinsm} the existence of 
superstring symmetries which naturally suppress the proton 
decay mediating operators was studied, while in ref. \cite{cus}
it was shown that the free fermionic string models occasionally
give rise to non--Abelian custodial symmetries, which forbid 
proton decay mediating operators to all orders of non--renormalizable
terms. These proposals, however, fall short of providing 
a satisfactory solution. The reason being that these proposals
are, in general, exclusive to the generation of light neutrino
masses through a see--saw mechanism. for example, the absence
of the $SO(10)$ 126 representation in string models necessitates
that the $SO(10)$ $U(1)_{Z^\prime}$ be broken at a high scale,
rather than at a low scale. Similarly, to date, the existence 
of the custodial non--Abelian symmetries seems to be exclusive
to the generation of a see--saw mass matrix. I also remark that
the presence of additional gauge bosons in non--realistic string models
as been noted in ref. \cite{cvetic}, as well as a suggestion that
the low energy data hints on the existence of an additional 
$Z^\prime$ with stringy characteristics \cite{langacker}.

The above discussion highlights both the importance and difficulty
of finding a robust and satisfactory solution to the proton stability
problem. The solution which is advocated in this paper is that
unification of gravity and the gauge interactions necessitates
the existence of an additional $U(1)$ symmetry, beyond the 
Standard Model, which remain unbroken down to low or intermediate
energy. Furthermore, the required $U(1)$ symmetry is not of
the type that arises in $SO(10)$ or $E_6$ GUTs.
Invariance under the extra $U(1)$ forbids the proton decay 
mediating operators, which can be generated only
after $U(1)_{Z^\prime}$ breaking. The magnitude 
of the proton decay mediating operators is therefore
proportional to the $U(1)_{Z^\prime}$ breaking scale,
$\Lambda_{Z^\prime}$ which is in turn constrained
by the proton lifetime limit. On the other hand, the type
of $U(1)$ that we consider here do not forbid quark, lepton and
seesaw mass terms. 

By studying the spectrum and symmetries of the string model
of ref. \cite{eu} Pati showed \cite{pati} that $U(1)$ symmetries
with the required properties do indeed exist in the string models. 
In this paper I examine whether the $U(1)$ symmetries can remain
unbroken down to low, or intermediate, energy scale. This is
achieved by examining if there exist supersymmetric flat directions
which preserve the specific $U(1)$ combinations, and hence
allow them to remain unbroken down to low, or intermediate energies. 
In the model of ref. \cite{eu} I show that, in fact, such flat directions
do not exist. I then study the same question in other models 
and show that in some examples the required symmetries cannot be
preserved by the flat directions, whereas in some cases they can.
Imposing the proton lifetime limits I estimate the scale of
$U(1)_{Z^\prime}$ breaking, $\Lambda_{Z^\prime}$. I show that
in the absence of large R-parity violation $\Lambda_{Z^\prime}$
is not constrained to be within the reach of forthcoming 
accelerator experiments, whereas if there exists substantial
R-parity violation, the $Z^\prime$ gauge boson is likely to 
be seen in forthcoming collider experiments.


\section{Gauge symmetries in free fermionic models}\label{stringmodels}

In this section I discuss the general structure of the realistic
free fermionic models, and of the additional $U(1)$ symmetries
that arise in these models. It is important to emphasize that the
free fermionic heterotic--string formulation yields a large number 
of three generation models, which possess an underlying $Z_2\times Z_2$
orbifold structure, and differ in their detailed phenomenological 
characteristics. It is therefore important, as elaborated below,
to extract the features of the models that are common to this 
large class of realistic models. 


The free fermionic models are constructed by specifying a set
of boundary conditions basis vectors and the one--loop
GSO projection coefficients \cite{fff}.
The basis vectors, $b_k$, span a finite  
additive group $\Xi=\sum_k{{n_k}{b_k}}$
where $n_k=0,\cdots,{{N_{z_k}}-1}$, with 
$N_{z_k}$ the smallest positive integer such that
$N_{z_k} b_k = \vec{0}$ (mod 2). 
The physical massless states in the Hilbert space of a given sector
$\alpha\in{\Xi}$, are obtained by acting on the vacuum with 
bosonic and fermionic operators and by
applying the generalized GSO projections. The $U(1)$
charges, $Q(f)$, with respect to the unbroken Cartan generators of the four 
dimensional gauge group, which are in one 
to one correspondence with the $U(1)$
currents ${f^*}f$ for each complex fermion f, are given by:
\beqn
{Q(f) = {1\over 2}\alpha(f) + F(f)},
\label{u1charges}
\eeqn
where $\alpha(f)$ is the boundary condition of the world--sheet fermion $f$
in the sector $\alpha$, and 
$F_\alpha(f)$ is a fermion number operator counting each mode of 
$f$ once (and if $f$ is complex, $f^*$ minus once). 
For periodic fermions,
$\alpha(f)=1$, the vacuum is a spinor in order to represent the Clifford
algebra of the corresponding zero modes. 
For each periodic complex fermion $f$
there are two degenerate vacua ${\vert +\rangle},{\vert -\rangle}$ , 
annihilated by the zero modes $f_0$ and
${{f_0}^*}$ and with fermion numbers  $F(f)=0,-1$, respectively. 

The four dimensional gauge group in the three generation
free fermionic models arises as follows. The models can 
in general be regarded as constructed in two stages.
The first stage consists of the NAHE set of boundary conditions basis
vectors, which is a set of five boundary condition basis vectors, 
$\{{\bf1},S,b_1,b_2,b_3\}$ \cite{nahe}. 
The gauge group after imposing the GSO projections induced
by the NAHE set basis vectors is $SO(10)\times SO(6)^3\times E_8$
with $N=1$ supersymmetry. The space--time vector bosons that generate
the gauge group arise from the Neveu--Schwarz sector and
from the sector ${\bf1}+b_1+b_2+b_3$. The Neveu--Schwarz sector
produces the generators of $SO(10)\times SO(6)^3\times SO(16)$.
The sector $\zeta\equiv{\bf1}+b_1+b_2+b_3$ produces the spinorial  
${\bf128}$
of $SO(16)$ and completes the hidden gauge group to $E_8$.
At the level of the NAHE set the sectors $b_1$, $b_2$ and $b_3$
produce 48 multiplets, 16 from each, in the $16$ 
representation of $SO(10)$. The states from the sectors $b_j$
 are singlets of the hidden $E_8$ gauge group and transform 
under the horizontal $SO(6)_j$ $(j=1,2,3)$ symmetries. This structure
is common to all the realistic free fermionic models. At this stage 
we anticipate that the $SO(10)$ group gives rise to the Standard Model 
group factors, whereas the $SO(6)^3$ groups may produce additional 
symmetries that can play a role in safeguarding the proton lifetime. 

The second stage of the free fermionic
basis construction consists of adding to the 
NAHE set three (or four) additional boundary condition basis vectors. 
These additional basis vectors reduce the number of generations
to three chiral generations, one from each of the sectors $b_1$,
$b_2$ and $b_3$, and simultaneously break the four dimensional
gauge group. The $SO(10)$ is broken to one of its subgroups
$SU(5)\times U(1)$, $SO(6)\times SO(4)$, $SU(3)\times SU(2)^2\times U(1)$
or $SU(3)\times SU(2)\times U(1)^2$.
Similarly, the hidden $E_8$ symmetry is broken to one of its
subgroups by the basis vectors which extend the NAHE set.
This hidden $E_8$ subgroup may, or may not, contain $U(1)$ factors
which are not enhanced to a non--Abelian symmetry. As
the Standard Model states are not charged with respect to these
$U(1)$ symmetries, they cannot play a role in suppressing the 
proton decay mediating operators, and are therefore not discussed
further here. On the other
hand, the flavor $SO(6)^3$ symmetries in the NAHE--based models
are always broken to flavor $U(1)$ symmetries, as the breaking
of these symmetries is correlated with the number of chiral
generations. Three such $U(1)_j$ symmetries are always obtained
in the NAHE based free fermionic models, from the subgroup
of the observable $E_8$, which is orthogonal to $SO(10)$.
These are produced by the world--sheet currents ${\bar\eta}{\bar\eta}^*$
($j=1,2,3$), which are part of the Cartan sub--algebra of the
observable $E_8$. Additional unbroken $U(1)$ symmetries, denoted
typically by $U(1)_j$ ($j=4,5,...$), arise by pairing two real
fermions from the sets $\{{\bar y}^{3,\cdots,6}\}$,
$\{{\bar y}^{1,2},{\bar\omega}^{5,6}\}$ and
$\{{\bar\omega}^{1,\cdots,4}\}$. The final observable gauge
group depends on the number of such pairings. 

Subsequent to constructing the basis vectors and extracting the massless
spectrum the analysis of the free fermionic models proceeds by
calculating the superpotential. 
The cubic and higher-order terms in the superpotential 
are obtained by evaluating the correlators
\beq
A_N\sim \langle V_1^fV_2^fV_3^b\cdots V_N\rangle,
\label{supterms}
\eeq
where $V_i^f$ $(V_i^b)$ are the fermionic (scalar) components
of the vertex operators, using the rules given in~\cite{kln}.
Generically, correlators of the form (\ref{supterms}) are of order
${\cal O} (g^{N-2})$, and hence of progressively higher orders
in the weak-coupling limit.
Typically, 
one of the $U(1)$ factors in the free-fermion models is anomalous,
and generates a Fayet--Ilioupolos term which breaks supersymmetry
at the Planck scale. The anomalous $U(1)$ is broken, and supersymmetry
is restored, by a non--trivial VEV for some scalar
field that is charged under the anomalous $U(1)$.
Since this field is in general also charged with respect
to the other anomaly-free $U(1)$ factors, some non-trivial
set of other fields must also get non--vanishing VEVs $\cal V$,
in order to ensure that the vacuum is supersymmetric.
Some of these fields will appear in the nonrenormalizable terms
(\ref{supterms}), leading to
effective operators of lower dimension. Their coefficients contain
factors of order ${\cal V} / M{\sim 1/10}$.
Typically the solution of the D-- and F--flatness
constraints break most or all of the horizontal $U(1)$ symmetries. 
The aim of this paper is to examine whether the $U(1)$, proton
safeguarding, symmetries can remain unbroken in the supersymmetric vacuum. 

\section{Proton decay and superstring $Z^\prime$s}


The proton decay mediating terms in a supersymmetric theory are the 
dimension four operators
\beq
\eta_1QUD+\eta_2UDD
\label{d4op}
\eeq
and the dimension five operators
\beq
QQQL~{\rm and}~UUDE
\eeq
where generation indices are suppressed, and
where $Q$, $L$ are the quark and lepton $SU(2)_L$ doublets and
$U$, $D$, are the two quark $SU(2)_L$ singlets, and $E$ is the
charged lepton $SU(2)$ singlet. 

In the realistic free fermionic models the dimension four 
operators are forbidden by the gauged $B-L$ symmetry. 
However, they are effectively induced after the spontaneous
breaking of the $B-L$ symmetry, from the terms that include the
neutral lepton $SU(2)$ singlet, $N$, which is the Standard Model
singlet field in the 16 representation of $SO(10)$,
\beq
QLDN~+~UDDN~.
\eeq
The VEV of $N$ then induces the effective dimension four 
operators with effective Yukawa couplings $\eta\sim
\langle N\rangle/M_{\rm string}$. The important point is that 
in the string models, in the absence of the 
126 representation of $SO(10)$, the $B-L$ symmetry
is necessarily broken at a high scale in order to suppress 
the left--handed neutrino masses. This breaking is induced
either by the VEV of the right handed neutrino $N$, or
by a combination of fields that effectively carry the
$B-L$ charge of the right handed neutrino. Thus, the dimension
four operators are in general induced at some order
of nonrenormalizable terms. While it is not impossible 
that the order will be sufficiently large so as to
sufficiently suppress the proton decay, it will clearly
be a property of a very specific point in the string moduli
space and not a very robust explanation for the proton lifetime.

The problem with proton decay is rather generic in string derived
models in which the Standard Model spectrum possess an underlying
$SO(10)$ embedding due to the quartic 16 operator that exist in
$SO(10)$. Thus, the same problem persists in flipped $SU(5)$ 
string models and in the Pati--Salam string models. In fact,
in these cases the problem is worse because in these cases 
the right--handed neutrino is necessarily used to break the 
GUT $SU(5)$ or $SU(4)$ symmetry. 

We expect therefore that in superstring models the gauged $B-L$ symmetry
cannot provide adequate protection for the proton lifetime. The basic 
claim of this paper, therefore, is that, in addition to the Standard Model
gauge group, there should exist an additional $U(1)_{Z^\prime}$ symmetry,
which  forbids the proton decay mediating operators, and remains unbroken
to intermediate or low energies. These operators can therefore
arise only from higher order nonrenormalizable terms in the
superpotential that contain fields, which are charged under $U(1)_{Z^\prime}$. 
On the other hand, the $U(1)_{Z^\prime}$ must be broken above the 
electroweak scale, as its associated gauge boson has not been observed
experimentally. Consequently, the magnitude of the 
proton decay mediating couplings are proportional to the
$U(1)_{Z^\prime}$ breaking scale. The proton lifetime limits then
impose an upper bound on the scale of $U(1)_{Z^\prime}$ breaking. 
In addition to the suppression induced by the $U(1)_{Z^\prime}$
breaking scale, the couplings may also be suppressed because of
the order at which they appear in superpotential. That is
the couplings may be forbidden by additional $U(1)$ symmetries
that are broken near the Planck scale and induce suppression factors
of order (1/10), as discussed in section \ref{stringmodels}.
The magnitude of the effective Yukawa couplings is affected
by the order of the nonrenormalizable terms that induce the effective
couplings. For example, in the string model of ref. \cite{eu}, we find that
the dimension four and five operators can arise from the order
sixth terms\footnote{for the notation 
and charges see ref. \cite{eu}},
\begin{eqnarray}
 &(u_3d_3+Q_3L_3)d_2N_2\Phi_{45}{\bar\Phi}_2^{-}\nonumber\\
+&(u_3d_3+Q_3L_3)d_1N_1\Phi_{45}\Phi_1^{+}\nonumber\\
+&u_3d_2d_2N_3\Phi_{45}{\bar\Phi}_2^{-}+
  u_3d_1d_1N_3\Phi_{45}\Phi_1^{+}\nonumber\\
+&Q_3L_1d_3N_1\Phi_{45}\Phi_3^+
+Q_3L_1d_1N_3\Phi_{45}\Phi_3^+\nonumber\\
+&Q_3L_2d_3N_2\Phi_{45}{\bar\Phi}_3^-
+Q_3L_2d_2N_3\Phi_{45}{\bar\Phi}_3^-.
\label{ordersix}
\end{eqnarray}
and 
\beq
Q_3Q_2Q_2L_3{\Phi_{45}}{\bar\Phi}_2^-~~~{\rm and}~~~
Q_3Q_1Q_1L_3{\Phi_{45}}{\Phi}_1^+
\label{n6terms}
\eeq
respectively, and additional terms are expected to arise 
at higher orders. Similar terms are found in the other
realistic free fermionic models. If we assume a GUT scale
VEV for $N$, and 1/10 suppression factors induced by the other
VEVs, we note that the effective dimension four and five operators
are not sufficiently suppressed, even if we consider generational mixing.

The question is then whether there exist string symmetries, which
are beyond the GUT symmetries and can provide an appealing explanation
for the proton lifetime. In a beautifully insightful paper 
\cite{pati} Pati studied this question
in the model of ref \cite{eu}, for the specific
choice of the $U(1)$ combinations that was given in \cite{eu}, and 
showed that such symmetries indeed exist in the string models.
The question that is studied here is whether the required symmetries
can in fact remain unbroken below the string scale, and hence
provide the needed suppression. 
The model of ref. \cite{eu}
contains six anomalous $U(1)$ symmetries:
${\rm Tr} U_1= {\rm Tr} U_2={\rm Tr} U_3=24,{\rm Tr} U_4= {\rm Tr} U_5=
{\rm Tr} U_6=-12$. These can be expressed by one anomalous
combination which is unique and five non--anomalous ones\footnote{
The normalization of the different $U(1)$ combinations is fixed
by the requirement that the conformal dimension of the
massless states still gives ${\bar h}=1$ in the new basis.}:
\beq
U_A={1\over{\sqrt{15}}}(2 (U_1+U_2+U_3) - (U_4+U_5+U_6))~;~ {\rm Tr} Q_A=
{1\over{\sqrt{15}}}180~.
\label{u1a}
\eeq
The choice for the five anomaly--free combinations
in ref. \cite{eu} is given by
\beqn
{U}_{12}&=& {1\over\sqrt{2}}(U_1-U_2){\hskip .5cm},{\hskip .5cm}
{U}_{\psi}={1\over\sqrt{6}}(U_1+U_2-2U_3),\label{u12upsi}\\
{U}_{45}&=&{1\over\sqrt{2}}(U_4-U_5){\hskip .5cm},{\hskip .5cm}
{U}_\zeta ={1\over\sqrt{6}}(U_4+U_5-2U_6),\label{u45uzeta}\\
{U}_\chi &=& {1\over{\sqrt{15}}}(U_1+U_2+U_3+2U_4+2U_5+2U_6).
\label{uchi}
\eeqn
The charges of the three generations,
$G_\alpha=E_{\alpha}+U_{\alpha}+N_{\alpha}+D_{\alpha}+
Q_\alpha+L_\alpha$ $(\alpha=1,\cdots,3)$, under the six unrotated 
$U(1)^{1,\cdots,6}$ are given below
\beqn
&({E}+{U})_{{1\over2},0,0,{1\over2},0,0}~+~
({D}+{N})_{{1\over2},0,0,{-{1\over2}},0,0}~+~
(L)_{{1\over2},0,0,{1\over2},0,0}~+~
(Q)_{{1\over2},0,0,-{1\over2},0,0}~,\\
&({E}+{U})_{0,{1\over2},0,0,{1\over2},0}~+~
({N}+{D})_{0,{1\over2},0,0,-{1\over2},0}~+~
(L)_{0,{1\over2},0,0,{1\over2},0}~+~
(Q)_{0,{1\over2},0,0,-{1\over2},0}~,\\
&({E}+{U})_{0,0,{1\over2},0,0,{1\over2}}~+~
({N}+{D})_{0,0,{1\over2},0,0,-{1\over2}}~+~
(L)_{0,0,{1\over2},0,0,{1\over2}}~+~
(Q)_{0,0,{1\over2},0,0,-{1\over2}}~.
\label{u1charges278}
\eeqn
where\footnote{$U(1)_C=3/2U(1)_{B-L}$; $U(1)_L=2U(1)_{T_{3_R}}$}
\beqn
{E}&&\equiv [(1,{3/2});(1,1)];{\hskip .6cm}
{U}\equiv [({\bar 3},-{1/2});(1,-1)];{\hskip .2cm}
Q\equiv [(3,{1/2});(2,0)]{\hskip 2cm}\\
{N}&&\equiv [(1,{3/2});(1,-1)];{\hskip .2cm}
{D}\equiv [({\bar 3},-{1/2});(1,1)];{\hskip .6cm}
L\equiv [(1,-{3/2});(2,0)]{\hskip 2cm}
\label{decomposition}
\eeqn
of $SU(3)_C\times U(1)_C\times SU(2)_L\times U(1)_L$.

$U(1)_\chi$ forbids the terms $UUDE$ and $LLEN$ but permits some 
$QLDN$, $UDDN$ and $QQQL$ terms. Therefore, if $U(1)_\chi$ remains unbroken 
down to low energies, it does not allow $R$-parity violation without 
inducing rapid proton decay. One must still insure that
the dimension four and five operator, which are allowed by $U(1)_\chi$
are sufficiently suppressed. 

$U(1)_\psi$ forbids all the proton decay mediating operators. Thus,
provided that $U(1)_\psi$ remains unbroken down to low energies,
the proton decay mediating operators may be sufficiently suppressed.
The viability of $U(1)_\psi$ as a symmetry which sufficiently
suppresses the proton decay mediating operators depends
on the $U(1)_\psi$ symmetry breaking scale. The required scale of
$U(1)_\psi$ breaking can be estimated by taking the 
relevant Yukawa couplings to be a function of the $U(1)_\psi$
breaking VEVs. On the other hand, $U(1)_\chi$ and $U(1)_\psi$
do not forbid the type of superpotential terms, $QU{\bar h}$, $QDh$,
$LEh$, $LN{\bar h}$ and $N{\bar N}\phi$, 
that generate the fermion masses, but may impose some restriction on
the textures of the of the fermion mass matrices. 

Examining Other phenomenological aspects of $U(1)_\psi$, we note that
$U(1)_\psi$  is family non--universal. General analysis of the fermion mass
matrices suggests that the states from the sectors $b_1$ and $b_2$
compose the heavy generation whereas $b_3$ gives rise to the light
generation \cite{nrt}. This means that the $U(1)_\psi$ combination produces
non--universal charges for the two light families. The existence
of a gauge boson with non--universal couplings for the two light
generations is constrained by Flavor Changing Neutral Currents
to be above 30 TeV. This problem, however, may be circumvented if
we redefine $U(1)_\psi$ as $2U_1-U_2-U_3$. With this redefinition
the superpotential terms leading to the dimension four
operators are still forbidden. However, assuming that the sector
$b_1$ produces the heavy generation and the sectors $b_2$ and
$b_3$ produce the two light generations, gives rise to universal
$U(1)_\psi$ charges for the two light generation, which are distinct from 
the heavy generation $U(1)_\psi$ charges. Thus, phenomenological
constraints on the viability of $U(1)_\psi$ at energy 
scales accessible to future experiments depend on detailed 
analysis of the fermion mass spectrum in the string models.

Next I turn to examine whether the symmetries $U(1)_\chi$ 
or $U(1)_\psi$ in the model of ref. \cite{eu} can remain
unbroken by the choices of F-- and D--flat directions.
To examine this question we extract the set of Standard 
Model singlets that are also neutral under $U(1)_\chi$
and $U(1)_\psi$. The set of fields which are neutral under
$U(1)_\chi$ contains $\{\Phi_{12},{\bar\Phi}_{12},
\Phi_{23},{\bar\Phi}_{23},\Phi_{13},{\bar\Phi}_{13}\}$,
and $T_i$, ${\bar T}_i$, which transform as $5$
and $\bar5$ of the hidden $SU(5)$ gauge group. Examining the
set of charges of these fields, it is seen that all
these fields are either neutral or carry positive charge
under the anomalous $U(1)_A$ symmetry. This means that at
least one field which is charged under $U(1)_\chi$ and 
carries negative charge under $U(1)_A$ must acquire a 
non--vanishing VEV in the cancellation of the anomalous
$U(1)_A$ D--term equation. Consequently, in the model
of ref. \cite{eu}, $U(1)_\chi$ is necessarily broken
by the supersymmetric flat directions, and cannot
play a useful role in suppressing the proton decay 
mediating operators. Similarly, the set of Standard Model
singlet fields which are neutral under $U(1)_\psi$ 
consist of $\{\Phi_{12},{\bar\Phi}_{12}\}$ and
$\{\Phi_{1,2,3}^\pm, {\bar\Phi}_{1,2,3}^\pm\}$. 
Again there is no solution to the D--term equations.
This results because the $\{\Phi_{1,2,3}^\pm, {\bar\Phi}_{1,2,3}^\pm\}$
states, which carry $Q_A=\pm1$ charges, also carry 
$Q_{2^\prime}=\mp2$ charges, whereas the 
$\{\Phi_{12},{\bar\Phi}_{12}\}$ states are neutral 
under both. Therefore, there cannot be a simultaneous 
solution for both $\langle D_A\rangle=0$ 
and $\langle D_{2^\prime}\rangle=0$. Therefore,
the two symmetries $U(1)_\psi$ and $U(1)_\chi$, in the model
of ref. \cite{eu}, cannot remain unbroken down to low energies
and cannot play a role in safeguarding the proton lifetime.
The possible reason for this result is that all the flat directions
that have been found in these model utilize the $SO(10)$ singlet field
$\Phi_{45}$, which seems to be necessary for D-flatness,
and is charged under $U(1)_\psi$ and $U(1)_\chi$.

One may contemplate the possibility in this model \cite{eu}
that a different choice of the anomaly free $U(1)$'s may produce a
$U(1)$ that forbids proton decay and can remain unbroken 
after implementing the F-- and D--flatness constraints.
Another choice of the anomaly free combinations is with
\beqn
{U}_{\psi^\prime} &=& {1\over\sqrt{21}}(3(U_1+U_2)-12 U_3 -4(U_4+U_5+U_6))\\
{U}_{\chi^\prime} &=& {1\over{\sqrt{210}}}(2(U_1+U_2)-U_3+2(U_4+U_5+U_6)).
\label{uchi2}
\eeqn
and the other combinations remain the same. The $U(1)_{\chi^\prime}$
symmetry now forbids all the proton decay mediating operators. 
However, in this case the only Standard Model singlets that are 
neutral under $U(1)_{\chi^\prime}$ are $\{N_{1,2}^c,\Phi_{12},
{\bar\Phi}_{12}\}$, which are either neutral, or carry positive
charge under the  the anomalous $U(1)_A$ symmetry. So again a solution for
the D--flatness constraints cannot exist with an unbroken
$U(1)_{\chi^\prime}$ and it cannot serve as the proton lifetime
safeguarding symmetry. 

The above discussion illustrates that despite the existence in the
string models of $U(1)$ symmetries that do forbid the proton decay 
mediating operators, it is not at all apparent that the needed
symmetry can remain unbroken below the string scale. This is in fact
a welcomed situation because it is seen that the string framework
is highly constrained. To exemplify this further we examine the
$U(1)$ symmetries in the FNY model of ref. \cite{fny}. 
In this model\footnote{The states and charges of the FNY model are given
in ref. \cite{fny,cfn}.}, prior to rotating the 
anomaly into a single $U(1)_A$, 
six of the FNY model's twelve $U(1)$ symmetries are anomalous:
Tr${\, U_1=-24}$, Tr${\, U_2=-30}$, Tr${\, U_3=18}$,
Tr${\, U_5=6}$, Tr${\, U_6=6}$ and  Tr${\, U_8=12}$.
Thus, the total anomaly can be rotated into a single 
$U(1)_{\rm A}$ defined by 
\beq
U_A\equiv -4U_1-5U_2+3U_3+U_5+U_6+2U_8.
\label{anomau1infny}
\eeq
The five orthogonal linear combinations,
\beqn
U^{'}_1 &=& \hbox to 3.0truecm{$2 U_1  - U_2 + U_3$\,\, ;\hfill}\quad 
U^{'}_2= -U_1 + 5 U_2 + 7 U_3\,\, ;\nonumber\\
U^{'}_3 &=& \hbox to 3.0truecm{$U_5 - U_6$\,\, ;\hfill}\quad 
U^{'}_4= U_5 + U_6 - U_8\,\, \label{nonau1}\\
U^{'}_5 &=& 12 U_1 + 15 U_2 - 9 U_3 + 25 U_5 + 50 U_8\,\, .
\nonumber
\eeqn
are all traceless
Note that in this case the anomalous $U(1)$ is not family universal. 
This arises because of the contribution to the anomaly of the ``Wilsonian''
sectors beyond the NAHE set. Therefore, it is not a priori apparent
that symmetries like the $U(1)_\chi$ and $U(1)_\psi$ can exist in
this model. Nevertheless, it is seen that in this model, for example,
$U^{'}_1$ forbids all the operators that can induce the dimension
four and five proton decay mediating operators. Furthermore, 
if we assume that the states from the sector $b_1$ form
the heavy generation, while those from $b_2$ and $b_3$
give rise to the two light generations, the charges of the 
two light generations are universal. Therefore, the $Z^\prime$
gauge boson associated with this symmetry is not strongly 
constrained by FCNC and could exist at energy scales accessible to
future colliders. Similarly, we find that the $U^{'}_2$ and  $U^{'}_5$
symmetries in this model forbid all the operators that can lead to proton
decay, whereas  $U^{'}_3$ and  $U^{'}_4$ and the unrotated $U_4$
do not. The two symmetries $U^{'}_2$ and  $U^{'}_5$ are 
family non--universal and therefore the associated $Z^\prime$s
are constrained to be heavier that $\sim30{\rm TeV}$.
A priori, unlike the case of the previous model,
it is not apparent that $U^{'}_2$ or  $U^{'}_5$
cannot survive the D--flatness constraints. However, a general
classification of the F-- and D--flat directions in the FNY model,
did not produce a vacuum in which either of those is preserved \cite{cfn}.
However, the vacua analyzed in ref. \cite{cfn} included stringent
flat directions, which imposes that they are flat to all orders
of non--renormalizable terms in the superpotential. Allowing
F--flatness breaking at a finite order may yield less
restrictive constraints. 

It is instructive to examine the same problem in the model
of ref. \cite{cus}. This model gives rise to custodial
symmetries \cite{symmetries} which forbid the proton decay
mediating operators to all orders of nonrenormalizable terms. 
However, as discussed above, the custodial symmetries, of the
type that arise in the model of ref. \cite{cus} may
be too restrictive and prevent application of the seesaw 
mechanism. The question that we want to explore is whether
this model can allow a symmetry like $U(1)_\psi$ to remain 
unbroken by the supersymmetric flat directions. The structure 
of this model is similar to that of ref. \cite{eu}. 
The model contains three anomalous $U(1)$ symmetries:
Tr${U_1}=24$, Tr${U_2}=24$, Tr${U_3}=24$. 
One combination remains anomalous and is 
given by: 
\beq
U_A=U_1+U_2+U_3,{\hskip 3cm}TrQ_A=72.
\label{u1a274}
\eeq
And the two orthogonal combinations can be taken as:
\beq
{U}_1^\prime=U_1-U_2{\hskip .5cm},{\hskip .5cm}
{U}_2^\prime=U_1+U_2-2U_3.
\label{afu1274}
\eeq
As in the model of ref. \cite{eu}, ${U}_2^\prime$ forbids the 
terms that can induce the proton decay mediating operators, whereas
${U}_1^\prime$ does not. However, it is again found that the model
does not admit flat directions that can leave ${U}_2^\prime$
unbroken at low energies. The reason being that in this model
$\Phi_{45}$, which is charged under ${U}_2^\prime$ must acquire
a non--vanishing VEV in the cancellation of the anomalous $U(1)_A$
D--term equation. 

The above discussion demonstrates that despite the fact that 
symmetries which forbid the proton decay mediating operators 
are abundant in the string models, it is not at all apparent
that they can remain unbroken down to low energies, and hence
fulfill the task of safeguarding the proton lifetime.

Similar results may be expected in the flipped $SU(5)$ \cite{fsu5}
and Pati--Salam type \cite{pssm} string models. These models
share the underlying $SO(10)$ structure, which is also 
possessed by the string standard--like models studied above.
The dimension four and five, proton decay mediating operators
arise from the quartic 16 operator in $SO(10)$ and are 
therefore common in all these models. In the flipped
$SU(5)$ the field assignment, in terms of $SU(5)\times U(1)$ 
representations, is $F=(10,1/2)\in\{Q,D,E\};~{\bar f}=
({\bar5},-3/2)\in\{U,L\};~{\rm and}~\ell^c=(1,5/2)\in\{E\}$.
The dimension four and
five, baryon and lepton number violating operators, then arise from 
$\{QQQL,QLDN,UDDN\}\rightarrow FFF{\bar f}$ and 
$\{UUDE,LLEN\}\rightarrow {\bar f}{\bar f}\ell^cF$. 
In the Pati--Salam type string models the field assignment,
in terms of the $SU(4)\times SU(2)_L\times SU(2)_R$
representations is: $F_L=(4,2,1)\in\{Q,L\};~
{\rm and} F_R=({\bar 4},1,2)\in\{U,D,E,N\}$. 
The proton decay mediating operators then arise from
$\{QQQL\}\rightarrow {F_LF_LF_LF_L}$, $\{UDDN,UUDE\}
\rightarrow {F_RF_RF_RF_R}$, and $\{QLDN,LLEN\}\rightarrow
{F_LF_LF_RF_R}$.
Furthermore, the existence of an anomalous $U(1)_A$, which
primarily arises from the breaking pattern of $E_6\times
SO(10)\times U(1)_A$, is also common to these models. 
Thus, it may be expected, although not proven, that
the symmetries like $U(1)_\psi$ above, are in general
broken near the string scale in this class of models. 

To show that indeed the required symmetries can in fact
survive down to low energies I turn to the left--right symmetric
models of ref. \cite{cfs}. The unique feature of these models,
in contrast to the standard--like, the flipped
$SU(5)$ and the Pati--Salam type string models, 
is that the anomalous $U(1)_A$ does not arise 
from the symmetry breaking pattern $E_6\times
SO(10)\times U(1)_A$ \cite{cfs}. First, I recap
the field theory content of these models. 
The observable sector gauge symmetry is 
$SU(3)_C\times SU(2)_L\times SU(2)_R\times U(1)_{B-L}$. 
Such models are reminiscent of the Pati--Salam
type string models, but differ from them by the fact that the $SU(4)$
gauge group is broken to $SU(3)\times U(1)_{B-L}$ already at the 
string level. Similar to the Pati--Salam models \cite{ps}, the left--right
symmetric models possess the $SO(10)$ embedding. 
The quarks and leptons are accommodated in the following
representations:
\beqn
Q_L^{i} &=& (3,2,1)_{1\over6}  ~~=~ {u\choose d}^i\\
Q_R^{i} &=& ({\bar 3},1,2)_{-{1\over6}} ~=~  {d^{c}\choose u^{c}}^i\\
L_L^{i} &=& (1,2,1)_{-{1\over2}} ~=~  {\nu\choose e}^i\\
L_R^{i} &=& (1,1,2)_{1\over2} ~~=~  {e^c\choose \nu^c}^i\\
h       &=& (1,2,2)_0 ~~=~  
{\left(\matrix{
                 h^u_+  &  h^d_0\cr
                 h^u_0  &  h^d_-\cr}\right)}
\label{LRsymreps}
\eeqn
where $h^d$ and $h^u$ are the two low energy supersymmetric superfields
associated with the Minimal Supersymmetric Standard Model. The breaking
of $SU(2)_R$ could be achieved with the VEV of $h$. However, this
will result with too light $W_R^\pm$ gauge boson masses. Additional
fields that can be used to break $SU(2)_R$ must therefore be postulated.
The simplest set would consist of two fields $H+\overline{H}$
transforming as $(1,1,2)_{-{1\over2}}+(1,1,{\bar 2})_{1\over2}$.
When $H$ and $\overline{H}$ acquire VEVs along their neutral
components $SU(2)_R\times U(1)_{B-L}$ is broken to the 
Standard Model weak--hypercharge, $U(1)_Y$.
The VEV of the Higgs multiplets $H+\overline{H}$
breaks the $B-L$ symmetry spontaneously and, in general,
induces the dimension four proton decay mediating 
operators, whereas the dimension five operators pose
a danger irrespective of this VEV. Thus, the need for
additional symmetries which suppress these terms is
again noted. In terms of the $SU(3)\times SU(2)_L\times SU(2)_R\times
U(1)_{B-L}$ representations, the baryon and lepton number
violating operators arise from
$\{QQQL\}\rightarrow {Q_LQ_LQ_LL_L}$, $\{UDDN,UUDE\}
\rightarrow {Q_RQ_RQ_RQ_R}$, $\{QLDN\}\rightarrow
{Q_LQ_RL_LL_R}$, and $\{LLEN\}\rightarrow L_LL_LL_RL_R$.
We can now examine, in the left--right symmetric string models
of ref. \cite{cfs}, whether the dangerous operators 
are still forbidden by symmetries like $U(1)_\psi$. 
The key feature of the left--right symmetric string models
which differs from the previous string models discussed above,
is the  $U(1)$ charge assignments of the three generation under
$U(1)_{1,2,3}$. 
In the flipped $SU(5)$, the Pati--Salam, and the standard--like, 
string models, the charges of a generation from a sector
$b_j$ $j=1,2,3$, under the corresponding symmetry $U(1)_j$ 
are either $+1/2$ or $-1/2$, for all the states from that sector.
In contrast, in the left--right symmetric string models the
corresponding charges, up to a sign are, 
\beq
Q_j(Q_L;L_L)=+1/2~~~;Q_j(Q_R;L_R)=-1/2,
\label{u1chargesinlrsmodel}
\eeq
{\it i.e.} the charges of the $SU(2)_L$ doublets have the opposite 
sign from those of the $SU(2)_R$ doublets. 
This is in fact the reason that in the left--right symmetric models
it was found that, in contrast to the case of the FSU5, PS and standard--like,
string models, the $U(1)_j$ symmetries are not part of the anomalous
$U(1)$ symmetry \cite{cfs}.
We then note, for example, that the $$U(1)_\zeta=U_1+U_2+U_3$$ 
combination forbids the dimension five operator $Q_LQ_LQ_LL_L$
and the operator $Q_RQ_RQ_RL_R$, which induces the effective 
dimension four operator $UDD\langle N\rangle/M_{\rm string}$,
while it allows the operator $Q_LQ_RL_LL_R$, which induces the
dimension four operator $QLD\langle N\rangle/M_{\rm string}$.
Similarly, the $U(1)_\psi=U_1+U_2-2U_3$, which was examined
in the case of the standard--like models above, forbid the 
$Q_LQ_LQ_LL_L$ and $Q_RQ_RQ_RL_R$ terms, while it allows the 
$Q_LQ_RL_LL_R$ operator.
Thus, in these models $U(1)_\zeta$, or $U(1)_\psi$,
can indeed suppress the 
proton decay amplitude, while it allows for R--parity violation. 
On the other hand, because $U(1)_j$ $(j=1,2,3)$ are anomaly free
the $U(1)_\zeta$, or $U(1)_\psi$, combinations
can remain unbroken down to low energies.
Furthermore, it is noted that the $U(1)$ combinations which
protects the proton longevity are not of a GUT origin, but 
of an intrinsic string origin. Thus, we have the exciting possibility that,
for example, R--parity violation may be accompanied with 
an additional $Z^\prime$ gauge boson of intrinsic stringy origin.
This demonstrates that the additional stringy $U(1)$ symmetries,
that play the role of safeguarding the proton lifetime,
can indeed remain unbroken down to low energies. 

\section{Estimate of the $Z^\prime$ mass}

The natural question that arises is at what scale can the
$U(1)$ symmetry, which protects the proton lifetime, be broken,
while still providing adequate suppression of the dangerous 
operators. This question, however, is rather model dependent
and depends on the order at which the nonrenormalizable terms,
which can induce the proton decay mediating operators, appear, 
and on possible additional suppression due to generational mixing.
Therefore, here I only attempt a rough estimate of the required
scale, in the case with and without R--parity violation. 
The dimension four operators that give rise to rapid proton decay,
$\eta_1UDD+\eta_2QLD$, are induced from the non--renormalizable 
terms of the form 
\beq
\eta_1(UDDN)\Phi+\eta_2(QLDN)\Phi^\prime
\eeq
where, $\Phi$ and $\Phi^\prime$ are combinations of fields 
that fix the  string selection rules. The field $N$ can be
the Standard Model singlet in the 16 representation of $SO(10)$, or
it can be a product of two fields, which effectively reproduces 
the $SO(10)$ charges of $N$ \cite{ps94}. I take the VEV of $N$, which 
breaks the $B-L$ symmetry, to be of the order of the GUT scale,
{\it i.e.} $\langle N\rangle\sim10^{16}{\rm GeV}$. This is required
because the VEV of $N$ induces the seesaw mechanism, which suppresses
the left--handed neutrino masses. The VEVs of $\Phi$ and $\Phi^\prime$
then fixes the magnitude of the effective proton decay mediating 
operators, with 
\beq
\eta_1^\prime\sim{{\langle{N}\rangle}\over{M}}
\left({\langle\phi\rangle^n\over{M^n}}\right)~~;~~
\eta_2^\prime\sim{{\langle{N}\rangle}\over{M}}
\left({\langle{\phi^\prime}\rangle^n\over{M^n}}\right).
\eeq
I take $M$ to be the heterotic string unification scale, which
is of order $10^{18}{\rm GeV}$. 
Similarly, the dimension five proton decay mediating operator
$QQQL$ can effectively be induced from the nonrenormalizable
terms 
\beq
\lambda_1 QQQL(\Phi^{\prime\prime})
\eeq
The VEV of $\phi^{\prime\prime}$ then fixes the
magnitude of the effective dimension five operator to be
\beq
\lambda_1^\prime\sim\lambda_1(
{{\langle\phi^{\prime\prime}\rangle^n}\over{M^n}}
\eeq
The experimental limits impose that the product 
$(\eta^\prime_1\eta_2^\prime)\le10^{-24}$ and
$(\lambda_1^\prime/M)\le 10^{-25}$. Hence, for 
$M\sim M_{\rm string}\sim 10^{18}{\rm GeV}$ 
we must have $\lambda_1^\prime\le10^{-7}$, to guarantee that
the proton lifetime is within the experimental bounds.
Assuming that the dimension four operators are induced at the
quintic order, {\it i.e.} with one additional field, that breaks
the proton protecting $U(1)_{Z^\prime}$ at intermediate energy
scale $\Lambda_{Z^\prime}$, we have 
\beq
({\eta^\prime_1\eta^\prime_2})\sim
\left({{\langle N\rangle}\over{M}}\right)^2
\left({{\Lambda_{Z^\prime}}\over{M}}\right)^2
\eeq
Taking $\langle N\rangle\sim 10^{16}{\rm GeV}$ and 
$M\sim 10^{18}{\rm GeV}$, 
we obtain the estimate $\Lambda_{Z^\prime}\le 10^8{\rm GeV}$.
similarly, from the dimension five operator we obtain the weaker
constraint $\Lambda_{Z^\prime}\le 10^{11}{\rm GeV}$. Thus, even
in the best case scenario $\Lambda_{Z^\prime}$ is not constrained to be
within the reach of forthcoming collider experiments. On the other hand,
if there exist sizable R--parity violation, which necessitates
one of the dimension four effective couplings, say $\eta_1^\prime$,
to be of order $O(10^{-5}-1)$, it imposes that the other effective
dimension four operator, say $\eta_2^\prime$, is of the order
$\eta_2^\prime\sim O(10^{-19}-10^{-24})$. Taking the smaller value
for $\eta_1^\prime$, and again taking
$\langle N\rangle\sim10^{16}{\rm GeV}$, this allows for the
$Z^\prime$ breaking scale to be as low as $10^{-17}M$, which is 
clearly too low. However, other suppression factors can arise
from generation mixing, and other VEVs which are of the order of
the Fayet--Iliopoulos D--term, and produce suppression of the 
order $\langle\phi\rangle/M\sim1/10$. These additional suppression
factors will result in elevating the $Z^\prime$ breaking scale
by two--four orders of magnitude. All in all, existence
of R--parity violating operator may well be accompanied by an
additional gauge boson of an intrinsic stringy origin.
The exciting prospect is to correlate
between R-parity violation and an additional $Z^\prime$ gauge boson,
whose properties depend on the particular string vacuum.

\section{Conclusions}

The structure of the Standard Model spectrum indicates the
realization of grand unification structures in nature. 
On the other hand the proton longevity severely constrains
the possible extensions of the Standard Model and
serves as a useful guide in attempts to understand the
origin of the Standard Model gauge and matter spectrum.
The realistic free fermionic heterotic--string models
reproduce the grand unification structures that are suggested
by the Standard Model and represent the most realistic
string models constructed to date. As such the realistic 
free fermionic models serve as a useful probe to the
fundamental characteristics of the possibly true 
string vacuum, as well as to various properties that
the string vacuum should possess in order to satisfy various
phenomenological constraints. In this paper, I proposed
that proton stability necessitates the existence of 
an additional $U(1)$ symmetry, which remains unbroken down to
intermediate or low energies. Furthermore, the required symmetry
is not of the type that arises in Grand Unified Theories,
but is of intrinsic string origin. The realistic free fermionic 
models do indeed give rise to $U(1)$ symmetries, which are
external to the GUT symmetries, and forbid the proton decay 
mediating operators. By studying the supersymmetric flat
direction I showed that in some cases the required symmetries
cannot remain unbroken in the supersymmetric vacuum,
whereas in others they can. Estimate of the $Z^\prime$
mass reveals that if R-parity violating operators with 
couplings in the range $O(10^{-5}-1)$ exist, then the associated
$Z^\prime$ is likely to be seen in forthcoming collider
experiments, whereas if the R--parity violating operators 
are much suppressed, the $Z^\prime$ is not constrained to
be in the accessible energy range. The phenomenology 
associated with the additional gauge bosons in the string
models will be reported in forthcoming publications.


\bigskip
\medskip
\leftline{\large\bf Acknowledgments}
\medskip

It is a pleasure to thank Sacha Davidson for useful discussions. 
This work was supported in part by a PPARC advanced fellowship.


\vfill\eject

\bibliographystyle{unsrt}

\end{document}